\documentstyle[prd,aps,floats,psfig]  {revtex}

\def\lsim{\mathrel{\lower2.5pt\vbox{\lineskip=0pt\baselineskip=0pt
\hbox{$<$}\hbox{$\sim$}}}}
\def\gsim{\mathrel{\lower2.5pt\vbox{\lineskip=0pt\baselineskip=0pt
\hbox{$>$}\hbox{$\sim$}}}}

\newcommand{\ima}{{\mbox{Im}\,}}
\newcommand{\rea}{{\mbox{Re}\,}}
\newcommand{\be}{\begin{equation}}
\newcommand{\ee}{\end{equation}}

\newcommand{\NP}[1]{Nucl.\ Phys.\ {#1}}

\newcommand{\PL}[1]{Phys.\ Lett.\ {#1}}

\newcommand{\PR}[1]{Phys.\ Rev.\ {#1}}
\newcommand{\PRL}[1]{Phys.\ Rev.\ Lett.\ {#1}}

\begin{document}
\draft
\title{\LARGE \bf Complete meson-meson scattering
within one loop in Chiral Perturbation Theory: Unitarization and resonances
\footnote{To appear in the Proceedings of the IX International Conference on Hadron Spectroscopy, HADRON2001, IHEP, Protvino, Russia, August 2001}}



\author{\Large Jos\'e R. Pel\'aez and A. G\'omez Nicola}

\address{ \large 
  {Departamento de F\'{\i}sica Te\'orica II, Universidad Complutense,
28040 Madrid, SPAIN}}
\maketitle

\begin{abstract}
We review our recent one-loop calculation of all the two meson
scattering amplitudes within SU(3) Chiral Perturbation Theory,
i.e. with  pions, kaons and etas. By unitarizing these amplitudes we are
able to generate dynamically the lightest resonances in meson-meson scattering.
We thus obtain a remarkable description of the meson-meson
scattering data right from threshold up to 1.2 GeV, in terms
of chiral parameters in good agreement with previous determinations.
\end{abstract}

\date{\today}

\maketitle


\vspace{.7cm}

 Chiral Perturbation Theory (ChPT) \cite{weinberg}
provides a remarkable description of the dynamics
of pions, kaons and the eta, which are the pseudo-Goldstone
bosons associated to the spontaneous $SU(3)_L\times SU(3)_R$ chiral symmetry 
breaking down to $SU(3)_{R+L}$. The ChPT Lagrangian
contains the most general terms compatible
with  the symmetry breaking pattern, organized in a derivative
and mass expansion (generically $p$), which, for the amplitudes 
becomes an expansion in powers of the external momenta and the
masses over a scale of ${\cal
O}$(1 GeV). The loop divergences appearing at a given order
can be absorbed by a finite number of constants 
in the Lagrangian to the same order.
Thus, order by order, the theory is finite and 
predictive. This approach has been very successful 
at low energies (less than 500 MeV). 
However, it has been shown  that 
applying a coupled channel generalization of
the Inverse Amplitude Method (IAM) \cite{Truong,IAM1}
one gets a
very good description of meson-meson scattering up
to 1.2 GeV, generating dynamically seven
 light resonances \cite{IAM2}: the $\rho$, $K^*$,
$f_0$, $a_0$, the octet $\phi$, the $\sigma$ and the $\kappa$.
The properties of the last two, and even their existence, are subject 
to intense debate within the hadron spectroscopy community 
(see these proceedings). Remarkably, 
the chiral amplitudes unitarized 
with the IAM generate poles associated to these
two wide structures \cite{IAM2}. In principle, since
 this method is built from the perturbative ChPT results,
it should respect the good low energy constraints.
However, not all the one loop ChPT meson-meson scattering amplitudes
were known. Indeed, only 
$\pi\pi\rightarrow\pi\pi$ \cite{Kpi}, $K\pi\rightarrow K\pi$
\cite{Kpi}, $\eta\pi\rightarrow\eta\pi$ \cite{Kpi} and 
$K^+K^-\rightarrow K^+K^-$, $K^+K^-\rightarrow K^0
\bar{K}^0$ \cite{JAPaco} were available in the literature. 
Therefore additional
approximations had to be done \cite{IAM2}, which spoiled partially
the low energy regime and did not allow for a direct
comparison with the standard ChPT parameters.

Very recently \cite{nos}, we have
completed the one-loop meson-meson scattering calculation. 
There are three new amplitudes:
$K\eta\rightarrow K\eta$, $\eta\eta\rightarrow\eta\eta$ and
$K\pi\rightarrow K\eta$, but we have recalculated
the other five amplitudes unifying the  notation, ensuring exact perturbative
unitarity and also correcting some misprints in the literature.
Next, we have applied the coupled channel IAM  to these amplitudes.
Our results allow for a direct comparison with the
standard low-energy chiral parameters, which we find in very good
agreement with previous determinations from low-energy data. 
The main differences with \cite{IAM2}
are: i) we consider the full calculation of all the one-loop
amplitudes in dimensional regularization, ii) we are able to describe
simultaneously  the low energy and the resonance regions, and iii) 
we pay special attention to the estimation of uncertainties.

The ${\cal O}(p^2)$ 
scattering amplitudes (low energy theorems) are obtained 
from the lowest order Lagrangian at tree level, whereas
the ${\cal O}(p^4)$ calculation has the following contributions:
First, the one-loop diagrams in Fig.1, which are divergent.
Second, the tree level graphs with the second order Lagrangian, 
which depend on the chiral parameters $L_i$, that absorb the previous
divergences through renormalization. In Table I, we list some 
determinations of the
$L_i^r(\mu)$ renormalized in the usual $\overline{MS}-1$
scheme of ChPT, so that they depend
on a scale  (except $L_3$ and $L_7$), customarily
chosen at $\mu=M_\rho$.

\begin{figure}
\centerline{\hbox{\psfig{file=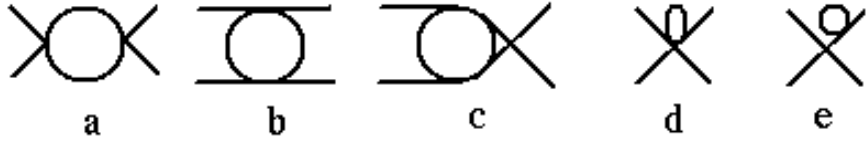,height=.08\textheight}}}

\vspace{.3cm}
  \caption{Generic one-loop Feynman diagrams that have to be evaluated
in meson-meson scattering.}
\label{fig1:diagrams}
\end{figure}

\begin{table}
\begin{tabular}{|c||c|c|c||c|}
\hline
 {Chiral Parameter} &
 {${\cal O}(p^6)$  $K_{l4}$ decays} &
 {${\cal O}(p^4)$  $K_{l4}$ decays} &
 {ChPT} &
 {IAM fits} \\
\hline
$L_1^r(M_\rho)$
& $0.53\pm0.25$
& $0.46$
& $0.4\pm0.3$& $0.56\pm0.10$ \\
$L_2^r(M_\rho)$
& $0.71\pm0.27$
& $1.49$
& $1.35\pm0.3$ & $1.21\pm0.10$\\
$L_3 $ & $-2.72\pm1.12$ & $-3.18$ & $-3.5\pm1.1$&
$-2.79\pm0.14$ \\
$L_4^r(M_\rho)$
& 0 (input)
& 0 (input)
& $-0.3\pm0.5$& $-0.36\pm0.17$ \\
$L_5^r(M_\rho)$
& $0.91\pm0.15$
& $1.46$
& $1.4\pm0.5$& $1.4\pm0.5$ \\
$L_6^r(M_\rho)$
& 0 (input)
& 0(input)
& $-0.2\pm0.3$& $0.07\pm0.08$ \\
$L_7 $ & $-0.32\pm0.15$ & $-0.49$ & $-0.4\pm0.2$&
$-0.44\pm0.15$ \\
$L_8^r(M_\rho)$
& $0.62\pm0.2$
& $1.00$
& $0.9\pm0.3$& $0.78\pm0.18$ \\
\hline
\end{tabular}
\caption{Several sets of chiral parameters ($\times10^{3}$) in the literature
as well as those from an IAM fit [7], with the
uncertainty due to different systematic error used on different
fits.} \label{eleschpt}
\end{table}

In order to compare with experiment,
we use partial waves $t_{IJ}$ of
definite isospin $I$ and angular momentum $J$. Thus, omitting 
the $I,J$ subindices, the chiral expansion becomes
$t\simeq t_2+t_4+...$, with $t_2$ and $t_4$ of ${\cal
O}(p^2)$ and ${\cal O}(p^4)$, respectively.
The unitarity relation  is
rather simple for the partial waves $t_{ij}$,
where $i,j$ denote the different available states. For
example, when two states, "1" and ``2'', are accessible, 
the partial waves satisfy
\be
\ima T = T \, \Sigma \, T^* \quad \Rightarrow \quad \ima T^{-1}=- \Sigma
\quad  \Rightarrow \quad T=(\rea T^{-1}- i \,\Sigma)^{-1}
\label{unimatrix}
\ee
with \vspace*{-.5cm}
\be
T=\left(
\begin{array}{cc}
t_{11}&t_{12}\\
t_{12}&t_{22} \\
\end{array}
\right)
\quad ,\quad
\Sigma=\left(
\begin{array}{cc}
\sigma_1&0\\
0 & \sigma_2\\
\end{array}
\right)\,,
\ee
where $\sigma_i=2 q_i/\sqrt{s}$ and $q_i$ is the C.M. momentum of
the state $i$. It can be readily noted that
{\it we only need to know the real part of the Inverse Amplitude}.
The imaginary part is fixed by unitarity.
Similar expressions can be obtained with $n$ accessible states. 
Note that, since the
unitarity relations are non-linear, they will never be satisfied 
exactly with a perturbative expansion like that of ChPT. Still,
unitarity holds perturbatively, i.e,
\vspace*{-.2cm}
\begin{eqnarray}
\ima T_2 = 0+ {\cal O}(p^4), \quad \quad \ima T_4 = T_2 \, \Sigma
\, T_2^*\,+ {\cal O}(p^6) . \label{pertuni}
\end{eqnarray}
A simple way to unitarize ChPT amplitudes is to
use in eq.(\ref{unimatrix})
the chiral expansion of 
$\rea T^{-1}\simeq  T_2^{-1}(1-(\rea T_4) T_2^{-1}+...)$.
Taking into account
 eq.(\ref{pertuni}),
we find
\begin{equation}
 T\simeq T_2 (T_2-T_4)^{-1} T_2,
\label{IAM}
\end{equation}
which is the coupled channel IAM, 
which we have used 
to unitarize simultaneously all the one-loop ChPT meson-meson
scattering amplitudes \cite{nos}.

Let us remark that since we have the
complete amplitudes renormalized in the $\overline{MS}-1$ scheme,
we can use  the IAM  with previous $L_i^r$ determinations. 
Still we find the correct resonant behavior.
Nevertheless we have carried out a fit (using MINUIT \cite{MINUIT})
of the available data on meson-meson scattering.  Since
there are incompatibilities between different experiments,
a $1\%$, $3\%$ and $5\%$ systematic error has been
added, which introduces an additional source of error. We give in
Table 1 the resulting chiral parameters from the fit, whose errors
correspond to those of MINUIT combined with those from the
systematic uncertainty. Note that they are compatible 
with previous determinations.
\begin{figure}
\hbox{\psfig{file=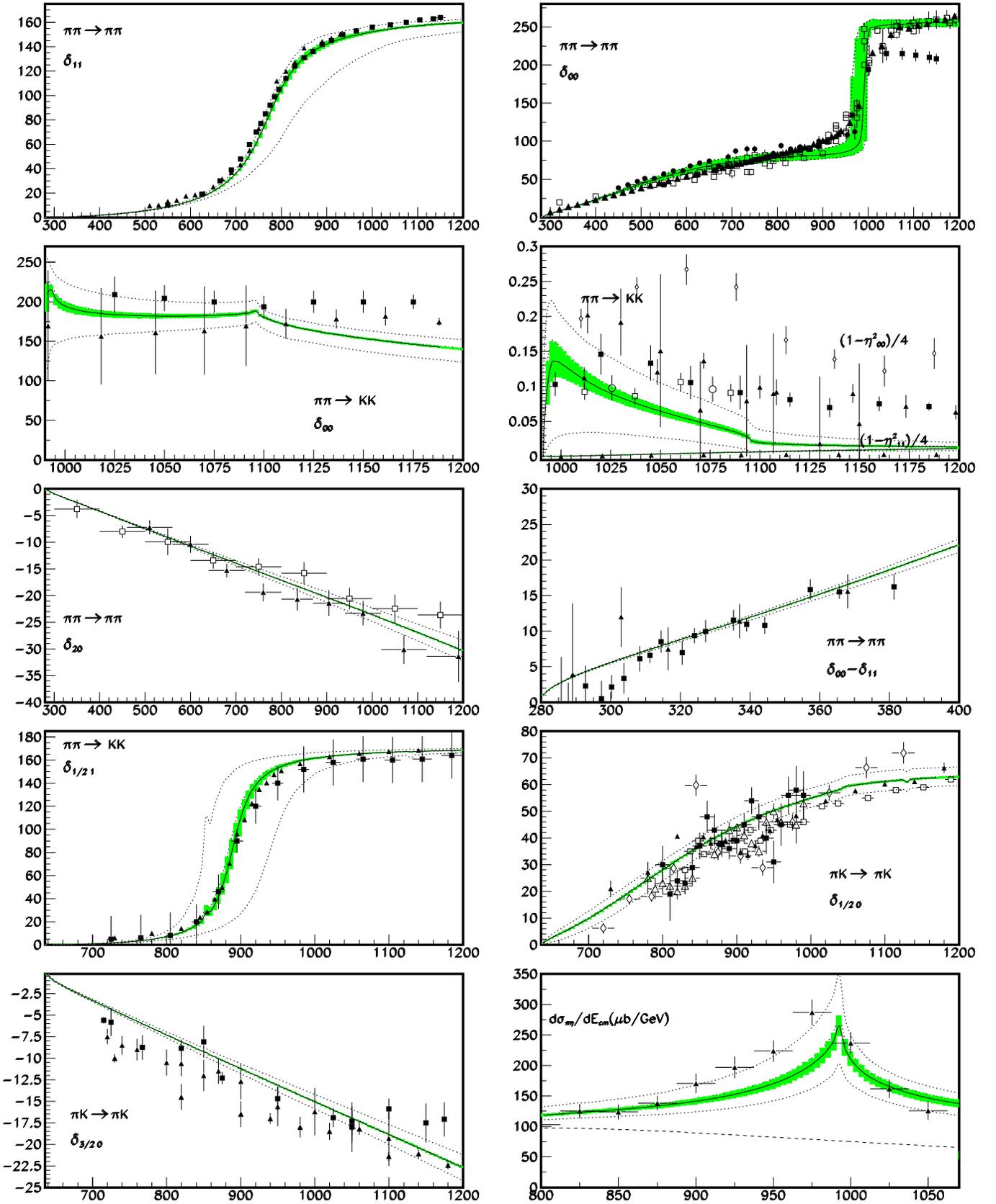,height=.95\textheight}}

\vspace{.3cm}
  \caption{Result of the coupled channel
IAM fit to meson-meson scattering data (see [7] for references).
The shaded area covers the uncertainty due to MINUIT errors.
The area between the dotted lines corresponds to the
uncertainty in the $L^r_i$  due to the
use of different systematic errors on the fits. 
The dashed line in the last plot is the continuous 
background underneath the resonant contribution.}
\end{figure}

In Fig.2 we show the IAM fit (see \cite{nos} for details).
The gray error bands cover the uncertainties in the $L_i$ 
due to MINUIT, and are obtained from a Monte-Carlo 
gaussian sampling of the parameters. Similarly, the area between the
dotted lines covers the errors
due to the different choice of systematic error.
Note that all the resonant features are
reproduced. However, since we have used the full one-loop
 amplitudes we are able
to obtain simultaneously values for the threshold parameters (they
have not been fitted) given in table 2. Note the good
agreement with the experimental values.

\begin{table}[h]
\begin{tabular}{|c|c|c|c|c|}
\hline 
Threshold&Experiment&IAM fit&ChPT ${\cal O}(p^4)$&ChPT ${\cal O}(p^6)$\\
parameter&&&\cite{explicitresonances,IAM1,Kpi}&\cite{bij}\\ 
\hline \hline 
$a_{0\,0}$&0.26 $\pm$0.05&0.231$^{+0.003}_{-0.006}$&0.20&0.219$\pm$0.005\\
$b_{0\,0}$&0.25 $\pm$0.03&0.30$\pm$ 0.01&0.26&0.279$\pm$0.011\\
$a_{2\,0}$&-0.028$\pm$0.012&-0.0411$^{+0.0009}_{-0.001}$&-0.042&-0.042$\pm$0.01\\
$b_{2\,0}$&-0.082$\pm$0.008&-0.074$\pm$0.001&-0.070&-0.0756$\pm$0.0021\\
$a_{1\,1}$&0.038$\pm$0.002&0.0377$\pm$0.0007&0.037&0.0378$\pm$0.0021\\ 
$a_{1/2\,0}$&0.13...0.24&0.11$^{+0.06}_{-0.09}$&0.17&\\
$a_{3/2\,0}$&-0.13...-0.05&-0.049$^{+0.002}_{-0.003}$&-0.5&\\
$a_{1/2\,1}$&0.017...0.018&0.016$\pm$0.002&0.014&\\
$a_{1\,0}$&&0.15$^{+0.07}_{-0.11}$&0.0072&\\ 
\hline
\end{tabular}
\vspace{.3cm}

\caption{ Meson-meson scattering lengths $a_{I\,J}$ and slope parameters
$b_{I\,J}$. For
experimental references see [7]. Note that our 
one-loop IAM results are very similar
to those of two-loop ChPT.}\label{elesfit}
\end{table}

\begin{paragraph}
{\bf Acknowledgments} Work  partially supported from the Spanish
CICYT projects FPA2000-0956, PB98-0782 and
BFM2000-1326.
\end{paragraph}




\begin{thebibliography}{99}
\vspace*{-.3cm}
\bibitem{weinberg} S. Weinberg, Physica A96, (1979) 327.
J. Gasser and H. Leutwyler, Ann. Phys. 158, (1984)
142 and  Nucl. Phys. B250,
(1985) 465,517,539.


\bibitem{Truong} T. N. Truong, \PRL{661}, (1988) 2526 ;\PRL{67}, (1991) 2260;
A. Dobado, M.J.Herrero and T.N. Truong, \PL{B235}, (1990) 134;

\bibitem{IAM1}
A. Dobado and J.R. Pel\'aez, \PR{D47}, (1993) 4883; \PR{D56}, (1997) 3057.
J. Nieves, M. Pav\'on Valderrama and E. Ruiz Arriola, hep-ph/0109077. 
\bibitem{IAM2}J. A. Oller, E. Oset and
J. R. Pel\'aez, Phys. Rev. Lett. 80, (1998)
3452; Phys. Rev. D59, (1999) 074001; Erratum-ibid. D60, (1999) 099906.

\bibitem{Kpi}
V. Bernard, N. Kaiser, U.G. Mei{\it ss}ner, \PR{D43} (1991), 2757;
\NP{B357} (1991), 129;  \PR{D44} (1991), 3698; \NP{B364} (1991), 283.

\bibitem{JAPaco}  F. Guerrero and J. A. Oller, Nucl. Phys. B537, (1999) 459.

\bibitem{nos} A. G\'omez Nicola and J. R. Pel\'aez, hep-ph/0109056.

\bibitem{BijnensKl4} G. Amor\'os, J. Bijnens and P. Talavera,
\NP{B602}(2001),87.
 Nucl.Phys.B585:293-352,2000, Erratum-ibid.B598:665-666,2001.


\bibitem{BijnensGasser} J. Bijnens, G. Colangelo and J. Gasser,
\NP{B427}, (1994) 427.

\bibitem{MINUIT}  F. James, Minuit Reference Manual D506 (1994).

\end{thebibliography}


\end{document}